# A Simple Descriptive Method & Standard for Comparing Pairs of Stacked Bar Graphs

© by Ronaldo Vigo, Ph.D.

*Center for the Advancement of Cognitive Science*

Ohio University

vigo@ohio.edu

ABSTRACT

While a plethora of research has been devoted to extoling the power and importance of data visualization, research on the effectiveness of data visualization methods from a human perceptual, and more generally, a cognitive standpoint remains largely untapped. Indeed, the way that human observers perceive and judge graphic charts can determine the interpretation of the graphed data. In this brief note we introduce a simple method for comparing stacked bar graphs based on a well-known result from cognitive science.

INTRODUCTION

A stacked bar graph (or stacked bar chart) is a chart that uses bars to show comparisons between categories of data as fractions of components of a whole. A bar represents a whole, and segments in the bar represent different parts or categories of that whole. For example, take the two stacked bar graphs in Figure 1 with their respective data tables. Upon visual inspection, the pair of stacked bar graphs on the left side of the figure are perceived to be of far greater similarity to each other than the pair on the right. Unfortunately, such visual assessment, as valuable as it may be, constitutes a subjective perceptual judgment. Many visualization techniques suffer from this potential criticism. Is human perception reliable enough to capture subtle qualitative and quantitative relationships by mere visual inspection? Regardless of the answer, methods should be available for the objective assessment of such perceptual judgments and their validity so that graphic reports can be expressed in terms of a referential standard whenever reliability ratings among judges are not available.

Such methods may be found in the form of successful mathematical models of cognition that were originally developed to predict and explain the way that humans perceive and conceptualize their physical and mental environments. We shall refer to the science of developing and applying such models from cognitive science to data science as *Cognitive Data Science*. Accordingly, in this brief note we apply two variants of a well-known cognitive model. The model in question was developed to predict human similarity judgments: thus, it may provide data scientists with an objective standard for evaluating the similarity between any pair of stacked bar graphs. Note that our claim is not that such model is capturing with precision human similarity judgments with respect to stacked graphs: the latter claim would have to be tested empirically. Instead, we argue that a well-known and successful measure of this type can be used as a measuring stick or standard for computing the pooled similarities between sets of pairwise comparisons with respect to stacked graphs and other similar graphic techniques that may be rendered or represented as vector quantities as will be shown. In what follows, we have purposely kept the mathematical details and rigor to a minimum in order to reach as wide an audience as possible. However, a more formal paper on this procedure is in progress.

THE PROBLEM

Often there are situations in data analysis when descriptive statistics reveal simple but useful patterns in the quantitative assessment of categorical data. A stacked bar chart is such a technique. Determining the degree of similarity and the overall similarity between pairs of stacked bar graphs is often useful when it is important to assess such similarity between many pairs and, in turn, take the average of such degrees of similarity. Regarding the latter, one possible scenario arises when comparing predictions from a model to the actual empirical data. One stacked bar graph may represent the model predictions while the other corresponds to the actual empirical data. Determining the average similarity scores from pairs of stacked bar graphs corresponding to different categories can then give a sense of how well the model can predict data arranged in terms of such categories.

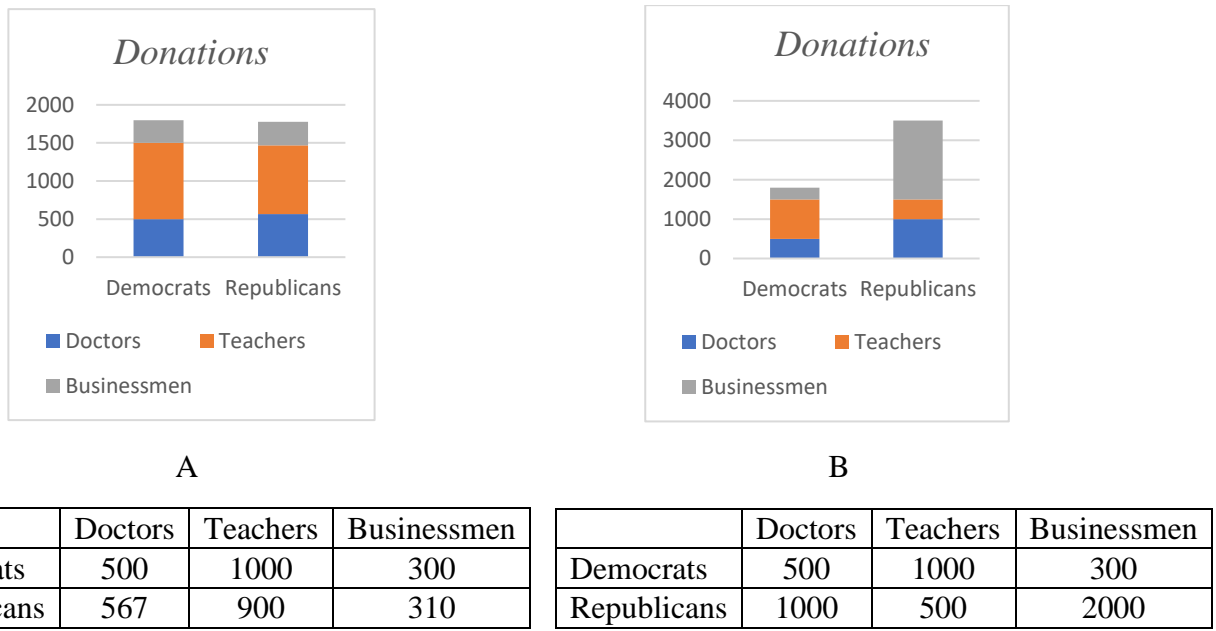

A

B

|  | Doctors | Teachers | Businessmen |
|---|---|---|---|
| Democrats | 500 | 1000 | 300 |
| Republicans | 567 | 900 | 310 |

|  | Doctors | Teachers | Businessmen |
|---|---|---|---|
| Democrats | 500 | 1000 | 300 |
| Republicans | 1000 | 500 | 2000 |

Figure 1. *On visual inspection, the pair of stacked column graphs on the left are clearly very similar to each other. The pair on the right are perceived as very dissimilar.*

To compute the similarity between stacked bar graphs one may use Shepard's similarity measure. We choose this measure because it has been a valuable tool in the field of human categorization research (Shepard, 1987; Nosofsky, 1984; Vigo, 2013; Vigo, 2015). I will use a simple version of this measure (i.e., without parameters) which states that the perceived degree of similarity between a pair of stimuli (also items, exemplars, objects) is inversely proportional to the exponent of the distance between the stimuli in psychological space. For a detailed explanation of the psychological theory, please see the referenced articles. The measure is shown in Equation 1.1.

Although various distance metrics may be used depending on the nature of the stimuli presented to subjects, for illustrative purpose we shall use only the Euclidean and Manhattan distance metrics. Let X and Y be vector representations of each object stimulus as follows: $X = (x_1, x_2, ..., x_n)$ and $Y = (y_1, y_2, ..., y_n)$ (where $X, Y \in \mathbb{R}^n$) so that $x_i$ and $y_i$ stand respectively for the value of the $i$-th dimension of stimulus X (or the $i$-th component of the vector X) and the value of the $i$-th dimension of stimulus Y (or the $i$-th component of vector Y). The generalized Minkowski distance $d(x, y)$ between the two stimuli X and Y is then defined by Equation 1.2 (when $r \geq 1$). As mentioned, in this note we shall use only the Euclidean distance ($r = 2$) as shown in Equation 1.3 because based on the aforementioned research it has been proven to work for integral dimensions (i.e., dimensions comprising the object-stimuli whose values are difficult to assess independently from the influence of the other, also integral, dimensions). On the other hand, the Manhattan metric (r=1) may be useful as well: particularly in situations involving clearly separable dimensions (see Equation 1.4).

The user should specify which metric is being used as a standard when apply the method introduced below. Also, note that when two points (or vectors) are identical in the metric space, the upper bound of the similarity measure computes to $e^{-0} = 1$; however, as the distance increases between stimuli, values decrease rapidly in an exponential fashion toward zero. Thus, the similarity value will lie somewhere in the [0,1] real number interval.

(1.1) $$s(X, Y) = e^{-d(X, Y)}$$

(1.2) $$d(X, Y) = \left[\left(\sum_{i=1}^{n} |x_i - y_i|^r\right)\right]^{1/r}$$

(1.3) $$d(X, Y) = \sqrt{\sum_{i=1}^{n} |x_i - y_i|^2}$$

(1.4) $$d(X, Y) = \sum_{i=1}^{n} |x_i - y_i|$$

To use the measure in Equation 1.1 with respect to stacked bar charts, we must first represent the stacked bar charts as vectors. In the case of Figure 1A above, we form two vectors: one with the components of the Republicans stacked bar chart and one with the components of the Democrats stacked bar chart. We can clearly do this because there exists a one to one mapping that preserves the order of the components of the two stacked bar charts. In fact, it is the aligned and consistent arrangement of the components of categories as data vectors in the first place that makes comparisons between stacked bar graphs meaningful. Let the Democrats stacked bar chart vector be X and the Republicans stacked bar chart vector be Y. We then get X= (500, 1000, 300) and Y= (567, 900, 310). Likewise, we can interpret the stacked bar graphs of Figure 1B as consisting of the vectors X= (500, 1000, 300) and Y= (1000, 500, 2000). We then rescale our vectors by dividing across by some constant c to get X= (500/c, 1000/c, 300/c) and Y= (1000/c, 500/c, 2000/c). This preserves all the proportions while facilitating the application of the exponential function. Indeed, c values that render the vector values in our comparisons to decimal values with at least one of the values being in the tenths are most convenient. As such, let c=1000 in our example. Researchers using the method should specify their chosen c value[1].

Thus, computing measure 1.1 using equation 1.2 we get that the similarity between the first pair of stacked bar charts is approximately .87 and for the second pair is approximately .16. As expected, these numbers correspond to our intuitions about the relative similarity between the two graphs. Other descriptive statistics such as the average similarity across a set or collection of pairs of stacked bar graphs (shown in Equation 1.5 below and as suggested in my modelling example above), their standard deviation, variance, etc. may also be computed according to the analytical needs of the researcher. In Equation 1.5, $s_i$ stands for the similarity (as computed above) between the $i$-th pair of stacked bar graphs (denoted $b_i$) in some collection $\omega = \{b_1, b_2, ..., b_m\}$ consisting of $m$ indexed pairs of stacked bar graphs.

In conclusion, these measures offer a standard for comparing stacked bar charts in situations when judges and reliability ratings from judges are not available and may provide a simple but uncluttered perspective on how well models fit data that have been arranged in terms of multiple categories.

(1.5) $$SIM_{avg}(\omega) = \frac{\sum_{i=1}^{m} s_i}{m}$$

---

[1] For the method described, we propose the following report format which includes the degree of similarity s, the type of distance metric r (r=1 for Manhattan or r=2 for Euclidean) and the scaling constant c for the pair of charts in question or for the vectors on which the charts are based: namely, (s=_, r=_, c=_).